\documentclass[a4paper,12pt]{article}
\usepackage[utf8]{inputenc}
\usepackage{amsmath,amssymb,amsthm,amscd,verbatim, graphicx, geometry, colortbl, xcolor, natbib}
\usepackage{thmtools}
\usepackage{thm-restate}
\usepackage{appendix}
\usepackage{textcomp}
\usepackage{booktabs}
\usepackage{multirow}
\usepackage{url}
\usepackage[flushleft]{threeparttable}
\usepackage{xcolor}
\newtheorem{remark}{Remark}
\newtheorem{theorem}{Theorem}
\newtheorem{corollary}{Corollary}

\title{On ratio measures of population heterogeneity for meta-analyses}
\author{Maxwell Cairns$^1$, Luke Prendergast$^2$ \\
La Trobe University\\
}

\begin{document}

\maketitle
\begin{abstract}
Popular measures of meta-analysis heterogeneity, such as $I^2$, cannot be considered measures of population heterogeneity since they are dependant on samples sizes within studies.  The coefficient of variation (CV) recently introduced and defined to be the standard deviation of the random effect divided by the absolute value of the overall mean effect does not suffer such shortcomings.  However, very large CV values can occur when the effect is small making interpretation difficult.  The purpose of this paper is two-fold.  Firstly, we consider variants of the CV that exist in the interval $(0,\ 1]$ making interpretation simpler.  Secondly, we provide interval estimators for the CV and its variants with excellent coverage properties.  We perform simulation studies based on simulated and real data sets and draw comparisons between the methods.  Based on our simulation and examples, we recommend transforming the CV onto this $(0,\ 1]$ domain for ease in interpretation and supported by a confidence interval estimate for this transformed variable.
\end{abstract}
\vspace{5cm}
Maxwell Cairns$^1$, Department of Mathematics and Statistics, La Trobe University, Melbourne, Australia; \\ \\
Luke Prendergast$^2$, Department of Mathematics and Statistics, La Trobe University, Melbourne, Australia.\\

Correspondence: Maxwell Cairns, Department of Mathematics and Statistics, La Trobe University, Melbourne, Australia 3086. Email: mrcairns994@gmail.com\\ \\
Luke Prendergast, Department of Mathematics and Statistics, La Trobe University, Melbourne, Australia 3086. Email: luke.prendergast@latrobe.edu.au
\newpage
\section{Introduction}
The presence of heterogeneity in a meta-analysis indicates that the true effects vary between studies. Also known as between-studies variance, this is a crucial part of any meta-analysis \citep{ThompsonSimonG.1994WSOH} since it is often plausible to assume that differences in studies (e.g. gender balance, average age etc.) lead to differences in the magnitude of the true effects. When heterogeneity is not assumed, that is, all effects are identical, a fixed-effect model (FEM) can be used. Let $Y_i$ be the estimator of the effect for the $i$th study, $\beta$ be the true effect to be estimated and $\epsilon_i \sim N(0,v_i)$ be the random sampling error where $v_i=\text{Var}(Y_i)$ is the random sampling error variance. Then, when there are $K$ studies under review for the meta-analysis, the FEM is of the form
\begin{equation}\label{eq:FEM}
    Y_i = \beta + \epsilon_i,\;\;(i=1,\ldots,K)
\end{equation}
where the $Y_i$s are all estimators of the same effect, $\beta$.

If a common effect for every study such as in the FEM is not plausible, then one option is to use the random-effects model (REM) which allows for variation between the individual study effects. This model has the form
\begin{equation}\label{eq:REM}
    Y_i = (\beta + \gamma_i) + \epsilon_i,\;\;(i=1,\ldots,K)
\end{equation}
where $\gamma_i \sim \text{N}(0, \tau^2)$ is the random effect and the estimated effect for the $i$th study is $(\beta + \gamma_i)$. When using an REM meta-analysis, understanding the extent of heterogeneity is important.  For example, even when $\beta$ is large, if $\tau$ is also large meaning that the true effects vary greatly, then it is possible that not all effects are clinically significant from a level indicating zero effect.

The structure of this paper is as follows.  Firstly we finish this introduction with a brief review of some popular measures of heterogeneity before providing a motivating example.  In Section 2 we discuss and compare several similar measures of heterogeneity. In Section 3 we provide the variance and bias of our variables. In Section 4 we introduce several methods for calculating confidence intervals for the three measures we discuss and review their performance via simulations in Section 5. In Section 6 we provide examples on how the three measures could be used in real meta-analyses, including returning to our motivating example before making some concluding comments.

\subsection{Inverse variance weighting and a popular measure of heterogeneity}

Let $W_i=1/v_i$ be the inverse variance weight (IVW) for the $i$th study.  The IVW estimator of the effect in FEM meta-analysis given as 
$$\widehat{\beta}=\sum^K_{k=i}w_iY_i$$ where $w_i = W_i/\sum^K_{i=1}W_i$ are the weights scaled to sum to one.  The IVW weights minimise the variance of $\widehat{\beta}$.  For a REM meta-analysis, the weights are $1/(v_i+\tau^2)$ which again minimise the variance of the estimator although in practice one must replace the unknown $\tau$ with its estimate.

For simplicity now and later, we use notation from \cite{biggerstaff1997incorporating} and let $S_r=\sum^K_{i=1}W_i^r$.  Cochrane's $Q$ \citep{cochran1954combination}, given as
$$Q= \sum^K_{i=1}W_iY_i- S_1^{-1}\sum^K_{i=i}(W_iY_i)^2,$$
is often used to facilitate tests for heterogeneity.  $Q$ also arises in the popular \cite{dersimonian1986meta} estimator of $\tau^2$ defined as
\begin{equation}
    T^2 = \text{max}\left\{0, \frac{Q-(K-1)}{S_1- S_2/S_1}\right\},\label{T2}
\end{equation}
where truncation is used at zero to avoid negative estimates of the random effect variance.

The most common statistic used to report levels of heterogeneity is $I^2$, which is the proportion of heterogeneity variance relative to the total variation.  It is given as
\begin{equation}
    I^2=\frac{\tau^2}{\tau^2 + \sigma_Y^2},\label{I2}
\end{equation}
where $\sigma^2_Y$ is chosen to be a typical within-study variance.  Proposed by \cite{higgins2002quantifying}, when the \cite{dersimonian1986meta} estimator of $\tau^2$ is used and and under their recommended choice for $\sigma^2_Y$, a common estimate of $I^2$ is simply

\begin{equation*}
    I^2=\frac{Q-df}{Q},
\end{equation*}
where $df=K-1$ is gthe degrees of freedom for a standard meta-analysis.  

An advantage of the measure is its simplicity. However, exactly what it is measuring is sometimes forgotten. As pointed out on page 118 of \cite{borenstein2011introduction}, $I^2$ is a descriptive statistic for inconsistency between the findings between studies, and not a measure of how much variation there is between the study effects. Hence, given that it is the proportion of variance explained by heterogeneity relative to the sum of the heterogeneity variance and within-study variance, a large estimated $I^2$ should not in itself lead to the conclusion that there exists a large amount of heterogeneity.  For example, if the within-study sample sizes are large, then the within-study variances can be very small leading to large $I^2$, even when $\tau$ is small.  Hence, if a large $I^2$ is used to highlight a potentially unreliable meta-analysis, then this needs to be done so by also considering the size of the $\tau$ estimate.  Misinterpretation of $I^2$ has received recent attention \citep[e.g.][]{rucker2008undue, hoaglin2016misunderstandings, kulinskaya2016commentary,JRSM:JRSM1230}.

\subsection{Motivating example}

We will now provide a motivating example based on a meta-analysis published in the \textit{Journal of Medical Virology} by \cite{zhu2020clinical}. The meta-analysis was performed on 35 studies using an REM and a double arcsin transformation of incidence rates. We have replicated the analysis using the \texttt{metafor} \citep{viechtbauer2010conducting} package in R \citep{R}. In their analysis, heterogeneity was assessed using the $Q$ statistic and $I^2$. 

\begin{figure}[h!t]
    \centering
    \includegraphics[scale = 0.6]{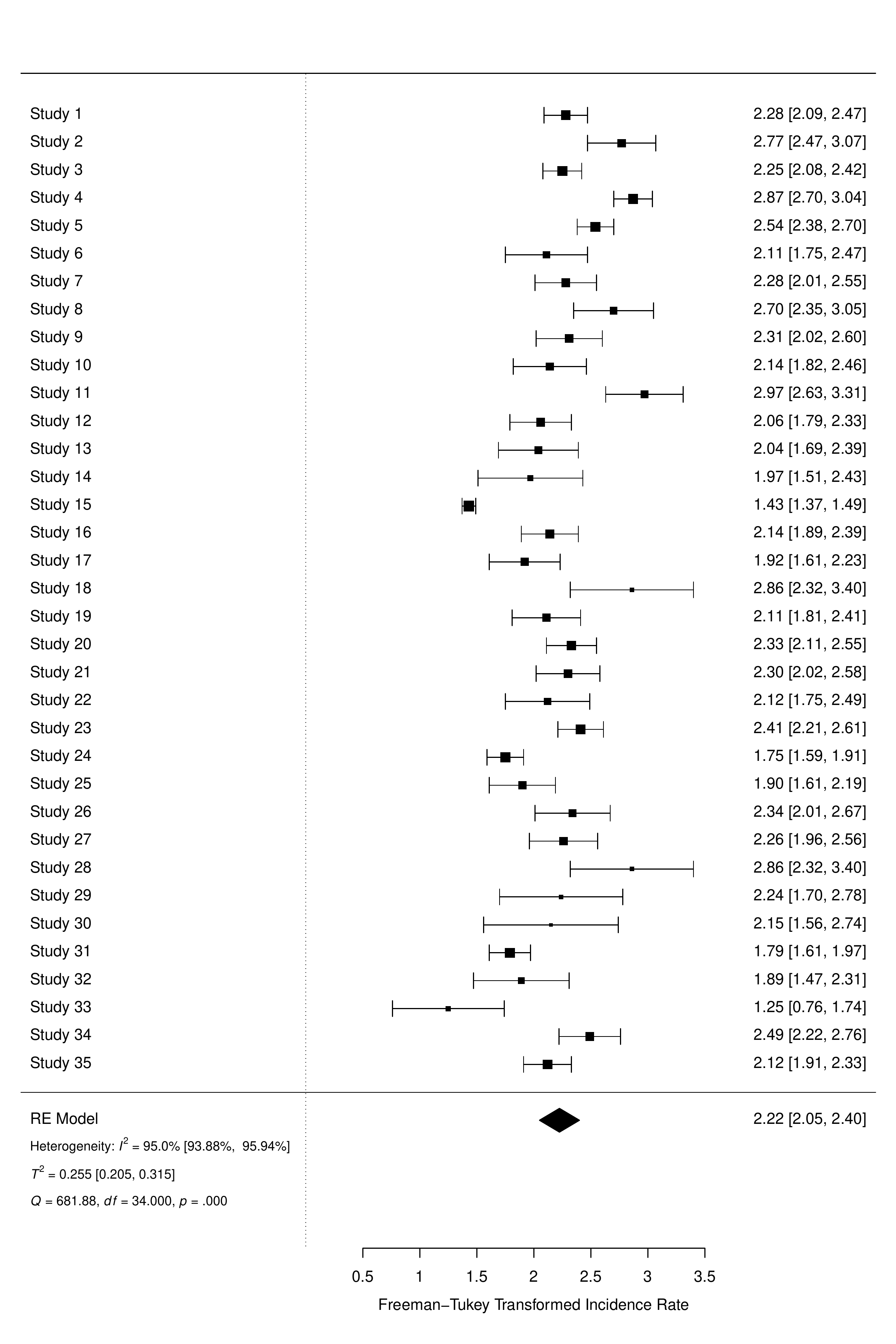}
    \caption{A replicated forest plot of Figure 3 in \cite{zhu2020clinical} which contains summary information of a meta-analysis containing 35 studies and heterogeneity statistics. Note that there are some minor discrepancies in the confidence intervals due to rounding.}
    \label{fig:mot_for}
\end{figure}

From the forest plot in Figure \ref{fig:mot_for}, we can see that $I^2$ is very large $(95\%)$ and that the $Q$ test indicates significant heterogeneity.   However, one should really be asking \textit{is the heterogeneity large in the context of what is being measured?}  One way to consider this is to consider the size of the estimated $\tau$ ($T$) relative to the estimated effect where we use $T$ instead of $T^2$ since $T$ and the estimated effect are measured on the same scale. We have $\sqrt{T^2}=\sqrt{0.255}\approx 0.5$, which is not large relative to the estimated effect size of 2.22.  In this context it is convenient then to use a measure of heterogeneity such as the coefficient of variation, denoted $\text{CV}_B$, introduced by \citep{takkouche1999evaluation, takkouche2013confidence}, defined to be the ratio of $T$ to the absolute value of the estimated effect.  The estimated CV$_B$ is 0.227 indicating towards small heterogeneity, which is in contrast to how $I^2$ is often interpreted \citep[e.g.][]{hoaglin2016misunderstandings}. 

\section{Some ratio measures of heterogeneity}\label{sec:heterogeneity}

We have previously discussed the use of $I^2$ and that it is often misinterpreted as a population measure of heterogeneity.  However, it is widely used because as a ratio measure it is independent of scale.  Below we list some properties we think are useful in the context of broader appeal for the use of measures of heterogeneity.  We consider measures to be the true values to be estimated, although for the properties below note that there estimator equivalents.

\begin{description}
\item[P1.] The measure is on an easily interpretable scale (e.g., [0,1]) or in the units of measurement of the effect.
\item[P2.] The measure can be interpreted as a population-level measure of heterogeneity.
\item[P3.] The measure can be applied to all random effects meta-analyses.
\item[P4.] Reliable interval estimators of the measure are available.
\label{d:P}
\end{description}

Unfortunately, popular measures for reporting heterogeneity do not have all of these properties. Most satisfy P1, but do not exhibit P2. Below we consider some other measures.

\subsection{Ratio of the random-effects and fixed-effect interval widths}
The Diamond Ratio \citep[DR,][]{cumming2016introduction}, so called because it is the ratio of the length of the REM diamond depicting the confidence interval for the mean effect to the length of FEM diamond, is one ratio measure of heterogeneity. Defined as
\begin{equation}
    DR = \frac{\sqrt{V_{RE}}}{\sqrt{V_{FE}}},
\end{equation}
where $V_{RE}$ and $V_{FE}$ are the estimated variances of the meta-estimator from the REM and FEM analyses. Originally proposed by \cite{higgins2002quantifying}, while like $I^2$ it is dependent on within-study sample sizes, it is less likely to be confused with a population measure of heterogeneity.  Instead, it can be viewed as a measure of trade-off in moving from the FEM analysis to the REM analysis.  This method allows the researcher a visual interpretation via forest plots, showing the percentage increase in the width of the REM confidence interval compared to the FEM confidence interval. Confidence intervals with very good coverage were recently proposed by \cite{cairns2020diamond}.

\subsection{Ratio of the heterogeneity standard deviation to the absolute effect}

One of the measures proposed by \cite{takkouche1999evaluation} for measuring heterogeneity was the between-study coefficient of variation given as
\begin{equation}
    \text{CV}_B=\frac{\tau}{|\beta|}.\end{equation}
In their paper the FEM estimator of $\beta$ was used in estimating CV$_B$, though this was changed in preference for the REM estimator by \cite{takkouche2013confidence} who also considered several confidence intervals for CV$_B$, namely Wald-type confidence intervals (i.e. estimate $\pm 1.96 \times \text{SE}$ where SE is the standard error) and bootstrap intervals.  An advantage of the CV$_B$ measure is that it does not depend on within-study sample sizes and can therefore be considered a population measure of heterogeneity. A potential drawback, is that very large values can be produced when $\beta$ is near zero. Hence, care needs to be taken when using measures with common effects such as standardized mean difference where an effect close to zero is not uncommon.

\subsection{Ratio of the heterogeneity variance to the effect estimator variance}
\cite{crippa2016new} introduced
\begin{equation}
    R_b = \frac{1}{K}\sum^K_{i=1}\frac{\tau^2}{v_i+\tau^2},\label{eq:Rb}
\end{equation}
which arises from $R_b=\tau^2/[K\text{Var}(\widehat{\beta})]$ where $\widehat{\beta}$ is the REM effect estimator. 
Clearly $R_b\in [0,1]$ and $R_b=1$ occurs when the within-study variances (variances of the study effect estimators) are zero, indicating maximum heterogeneity relative to within-study sampling variances.  Like $I^2$ and the DR, $R_b$ depends on the within-study variances and is therefore not a population measure of heterogeneity.  However, it was intended as a measure of heterogeneity between studies and should be interpreted as such.

\subsection{Ratio of the heterogeneity variance to the sum of the squared effect and heterogeneity variance} 

Another possibility is to adjust $\text{CV}_B$ so that it is valid when $|\beta|=0$, easier to interpret when $|\beta|$ is small, and still a measure of heterogeneity relative to the effect size.  As two possibilities we propose
\begin{equation}
    M_1 = \frac{\tau}{\tau + |\beta|}
\;\;\text{and}\;\;
    M_2 = \frac{\tau^2}{\tau^2 + \beta^2}.
\end{equation}

An advantage of these measures is that they are bounded in $[0,1]$ and therefore are simple to interpret.  In both cases, $M_i=1$ results when $\tau>0$ and $\beta=0$ so that heterogeneity variance is maximised relative to mean effect.   We note the link between $M_1$, $M_2$ and CV$_B$ where, by using the fact that $M_1^{-1} = 1 + 1/\text{CV}_B$ and $M_2^{-1}=1 + 1/\text{CV}_B^2$, we can write
\begin{equation}
    M_1=\frac{\text{CV}_B}{1 + \text{CV}_B}\;\;\text{and}\;\;M_2=\frac{\text{CV}^2_B}{1 + \text{CV}^2_B}.
    \label{E:CVlink}
\end{equation}

There are further links between CV$_B$, $M_1$ and $M_2$.  For example, consider the logit transformation defined to be, for a $u\in(0, 1)$, logit$(u)=\log[u/(1-u)]$ then we have that
\begin{equation}\label{links_between_measures}
    \text{logit}(M_1)=\log(\text{CV}_B)\;\;\text{and}\;\;\text{logit}(M_2)=2\log(\text{CV}_B).
\end{equation}

Note that \cite{takkouche2013confidence} produces confidence intervals for the CV$_B$ measure, and one of those is for the log transformed CV$_B$.  Hence, one possibility is to transform the interval for the $\log(\text{CV}_B)$ using the inverse logit transformation to obtain confidence intervals for $M_1$ and $M_2$.

\subsection{Comparisons}

Basic comparisons of the measures are of interest. We looked to see how many of the desired properties each measure satisfies (refer to Section \ref{d:P} for definitions). 
In Table \ref{T:compT} we summarise the measures considered in regards to satisfying properties P1-P4.  As to whether the measures are all applicable for every meta-analysis, work is needed in the context of meta-regression and we comment on this below in Remark \ref{remark:regression}.  \cite{takkouche2013confidence} proposed bootstrap and Wald-type intervals for the $R_b$ and CV$_B$ measures with the Wald intervals being the better performed overall.  However, while good coverage was achieved for many settings, in others the coverage was not close to nominal and often this was due to being too over-conservative.  We verify this later with our own simulations using the Wald intervals.   However, we also propose intervals for CV$_B$, $M_1$ and $M_2$ that do exhibit excellent coverages across our simulation studies.  For $R_b$, since it is a function of $\tau$ and the fixed within-study variances, good coverages should be possible using the a substitution approach with the $\tau$ confidence intervals.  \cite{cairns2020diamond} achieved good coverages using this approach for the DR which also depends on $\tau$ and within-study variances. 

\begin{table}[h!t]
\centering
\begin{threeparttable}
\caption{Properties P1-P4 satisfied (Y = Yes, N = No, ? = see Remark 1), for $I^2$, CV$_B$, DR, $R_b$, $M_1$ and $M_2$.}\label{T:compT}
\begin{tabular}{llllll} 
\toprule
  Measure   & P1 & P2 & P3 & P4 & Comments  \\ \midrule
  $I^2$ & Y & N & Y & Y & \\
  CV$_B$& N & Y & ? & Y$^*$ & See Remark 1 for P3 in regards to meta-regression \\
  DR & Y & N & ? & Y & See Remark 1 for P3  in regards to meta-regression\\
  $R_b$ & Y & N & Y & ? & Intervals available and others possible\\
  $M_1$ & Y & Y & ? & Y$^*$ & See Remark 1 for P3  in regards to meta-regression\\
  $M_2$ & Y & Y & ? & Y$^*$ & See Remark 1 for P3  in regards to meta-regression\\ \bottomrule
\end{tabular}
 \begin{tablenotes}
\item \footnotesize $^*$ Confidence intervals with excellent coverage for CV$_B$, $M_1$ and $M_2$ are proposed and assessed later.  
 \end{tablenotes}
\end{threeparttable}
\end{table}

\begin{remark}\label{remark:regression}
While $I^2$ and $R_b$ are naturally defined in a meta-regression analysis, this is not true for the CV$_B$, $DR$ and $M_1$ and $M_2$ since they vary depending on the level chosen for the moderator.  Possibilities exists, and this is part of some ongoing work, when a moderator is categorical in that we have a heterogeneity measures for each level of the moderator, and even for numeric moderators where the heterogeneity measure may make sense given a suitable reference points (e.g. for the moderator set to zero or the mean of all moderator values).  This ongoing work includes the construction of confidence intervals.
\end{remark}

Since they are measures of population heterogeneity, we will focus out attention throughout the rest of this paper on the CV$_B$, $M_1$ and $M_2$ measures, but in the below compare them to the most commonly reported heterogeneity measure, $I^2$.

\setlength{\tabcolsep}{3pt}
\begin{table}[h!t]
\centering
\caption{Summary statistics for 1000 simulated values for $I^2$, CV$_B$, $M_1$ and $M_2$ for varying $\beta$ and $\tau$.  The minimum (Min), first quartile ($Q_1$), median ($m$), third quartile $(Q_3)$ and maximum (Max) are reported.}\label{tab:sim_measures}
\begin{small}
\begin{tabular}{rrrrrrrrrrrrrrrrr}
  \hline
  & \multicolumn{5}{c}{$\tau=0$} & \multicolumn{5}{c}{$\tau=0.4$} & \multicolumn{5}{c}{$\tau=0.8$} \\
$\beta$ & Meas. & Min & $Q_1$ & $m$ & $Q_3$  & Max & Min & $Q_1$  & $m$ & $Q_3$  & Max & Min & $Q_1$  & $m$ & $Q_3$  & Max \\ 
  \hline
\multirow{4}{*}{$0.2$} & $I^2$ & 0 & 0 & 0 & 20.53 & 70.91 & 0 & 39.76 & 57.34 & 68.05 & 85.23 & 0 & 77.03 & 83.31 & 87.42 & 93.70 \\ 
 & CV$_B$  & 0 & 0 & 0 & 0.83 & 35.48 & 0 & 1.00 & 1.72 & 3.47 & 297.97 & 0 & 1.77 & 3.05 & 6.45 & 4983.89 \\ 
 & $M_1$  & 0 & 0 & 0 & 0.45 & 0.97 & 0 & 0.50 & 0.63 & 0.78 & 1.00 & 0 & 0.64 & 0.75 & 0.87 & 1.00 \\ 
 & $M_2$  & 0 & 0 & 0 & 0.41 & 1.00 & 0 & 0.50 & 0.75 & 0.92 & 1.00 & 0 & 0.76 & 0.90 & 0.98 & 1.00 \\ 
  \multirow{4}{*}{$0.5$} & $I^2$ & 0 & 0 & 0 & 21.76 & 69.35 & 0 & 40.21 & 57.84 & 68.77 & 85.67 & 7.80 & 76.25 & 82.74 & 86.91 & 93.91 \\ 
  & CV$_B$ & 0 & 0 & 0 & 0.35 & 1.45 & 0 & 0.50 & 0.77 & 1.09 & 80.77 & 0.17 & 1.00 & 1.39 & 2.21 & 12895.37 \\ 
 & $M_1$  & 0 & 0 & 0 & 0.26 & 0.59 & 0 & 0.34 & 0.43 & 0.52 & 0.99 & 0.14 & 0.50 & 0.58 & 0.69 & 1.00 \\ 
 & $M_2$  & 0 & 0 & 0 & 0.11 & 0.68 & 0 & 0.20 & 0.37 & 0.54 & 1.00 & 0.03 & 0.50 & 0.66 & 0.83 & 1.00 \\ 
  \multirow{4}{*}{$0.8$} & $I^2$ & 0 & 0 & 0 & 23.48 & 68.72 & 0 & 36.05 & 54.57 & 67.94 & 87.06 & 12.32 & 75.71 & 82.36 & 86.60 & 93.75 \\ 
 & CV$_B$  & 0 & 0 & 0 & 0.23 & 0.65 & 0 & 0.31 & 0.45 & 0.63 & 1.55 & 0.14 & 0.68 & 0.91 & 1.27 & 6.68 \\ 
& $M_1$  & 0 & 0 & 0 & 0.19 & 0.39 & 0 & 0.24 & 0.31 & 0.38 & 0.61 & 0.12 & 0.41 & 0.48 & 0.56 & 0.87 \\ 
 & $M_2$  & 0 & 0 & 0 & 0.05 & 0.30 & 0 & 0.09 & 0.17 & 0.28 & 0.71 & 0.02 & 0.32 & 0.45 & 0.62 & 0.98 \\ 
   \hline
\end{tabular}
\end{small}
\end{table}

\begin{figure}[h!t]
    \centering
    \includegraphics[scale=0.6]{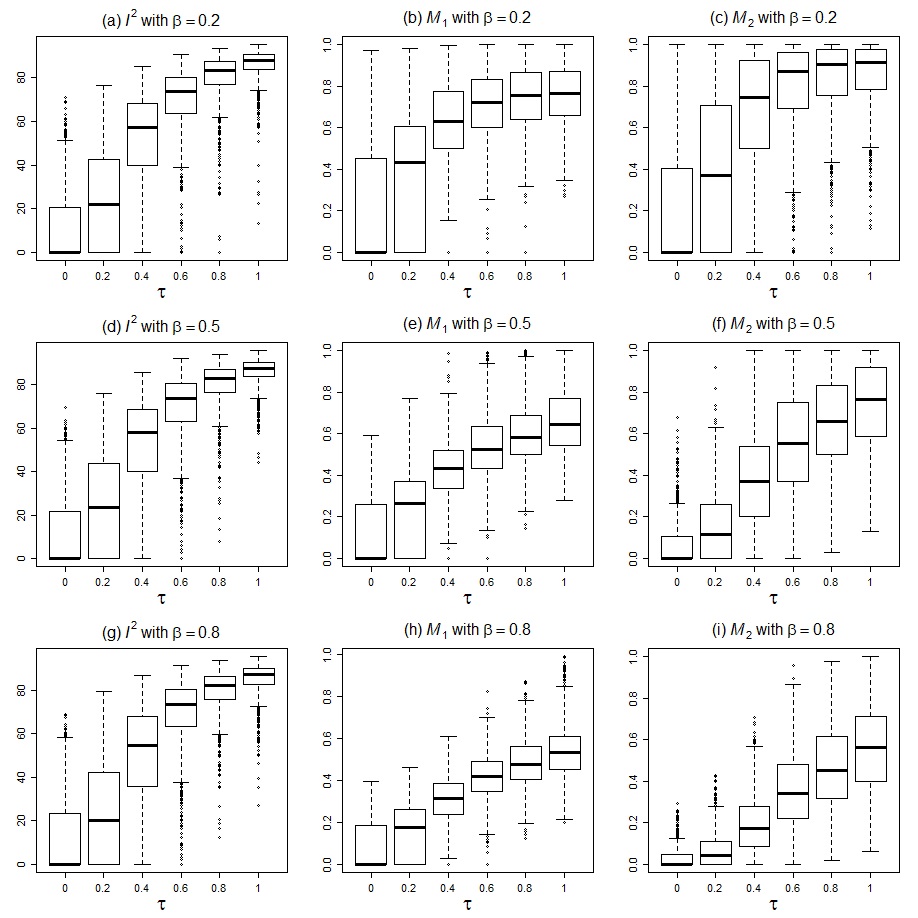}
    \caption{Boxplots of 1000 simulated values of $I^2$, $M_1$ and $M_2$ arising from randomly generated SMDs for varying $\tau$ (horizontal axis of each plot) and varying $\beta$ (over rows of plots).}
    \label{fig:my_label}
\end{figure}

We generated 1000 data sets consisting of 10 studies which two arms, each containing 10 observations.  We randomly sampled standardised mean differences (SMDs), see further details later, and conducted a meta-analysis on the generated data using the \texttt{rma} function from the R \texttt{metafor} package \cite{viechtbauer2010conducting}.  We used the DerSimonian and Laird estimator of $\tau$ \citep[DL,][]{dersimonian1986meta} since it most common across various packages and usually used to calculate $I^2$.  We varied the true effect over $0.2,0.5$ and $0.8$ (corresponding to small, moderate and large effects), with three choices of $\tau$, $0$ (zero heterogeneity), $0.4$ and 0.8.  We provide summary statistics of the simulations in Table \ref{tab:sim_measures}.  When there is zero heterogeneity, the median for all measures is zero.  As $\tau$ increases, so too does the median measure and the minimum of zero even when $\tau=0.8$ is due to a small number of trials where the DL estimator is truncated to zero.  However, the first quartiles are indicative of much higher heterogeneity.  The maximum for the CV$_B$ is sometimes very large which occurs due to small estimates of $\beta$.  When this happens $M_1$ and $M_2$ can be very close to one.  We also note that $I^2$ does not vary much at all with varying $\beta$, which is expected given that it compares the heterogeneity variance with within-study variance.  Hence, the measure gives similar results when $\tau=0.4$ regardless of the magnitude of $\beta$.

Figure \ref{fig:my_label} contains boxplots comparing the spread of $I^2$, $M_1$ and $M_2$ for the choices of $\beta$ in Table \ref{tab:sim_measures} with some additional choices for $\tau$.  We do not include CV$_B$ due to the sometimes large scale.  As expected and noted before, we see that there is not much change in the location and spread of $I^2$ as the value of $\beta$ increases.  However, for $M_1$ and $M_2$ values become smaller as $\beta$ increases indicating smaller relative heterogeneity in the context of effect size. 

\section{Point estimators}\label{sec:point}

As a ratio bounded on $(0, 1]$, point and interval estimation may be improved by transforming the estimators of $M_1$ and $M_2$ so that the transformed variables are on the domain $(-\infty,\ \infty)$.  We choose the logit transformation given as logit$(u/(1-u))$ for $u\in (0,\ 1)$ where back-transformation to the original scale can be achieved using its inverse, logit$^{-1}(v)=\exp(v)/(1 + \exp(v))$.  There are two notable advantages to this approach; firstly the bounds of interval estimators that are firstly computed on the logit scale and then back-transformed, are also bounded on $[0,\ 1]$.  Second, as noted previously in \eqref{links_between_measures} there is a strong link between transformed CV$_B$, $M_1$ and $M_2$ in that $\log(\text{CV}_B)=\text{logit}(M_1)=\text{logit}(M_2)/2$.

We now consider variances and bias for further investigation. We approximate the variance through use of the delta method (see Appendix A for a full proof).

\begin{theorem}\label{thm:var}
An approximate variance and bias for the logit transformed $M_1$, logit$(\widehat{M}_1)$, estimator are
\begin{align*}
        \text{Var}\left[\text{logit}(\widehat{M}_1)\right]&\approx \text{Var}(\widehat{\tau}^2)\cdot \frac{1}{4\tau^4} + \text{Var}(\widehat{\beta})\cdot \frac{1}{\beta^2},\\
    \text{bias}[\text{logit}(\widehat{M}_1)] &\approx\frac{1}{2}\left[\text{Var}(\widehat{\beta})\cdot\frac{1}{\beta^2}-\text{Var}(\widehat{\tau}^2)\cdot\frac{1}{2\tau^4}\right]
\end{align*}
where bias$(\cdots)$ denotes the bias of its estimator argument.
\end{theorem}

\cite{takkouche2013confidence} computed an approximate variance for the CV$_B$ estimator also using the delta method and asymptotic independence between the $\beta$ and $\tau$ estimators.  In noting that $\log(\widehat{\text{CV}}_B)=\text{logit}(\widehat{M}_1)$ as above, another way to also arrive at the variance in Theorem \ref{thm:var} is by using the variance for $\widehat{CV}_B$ and applying the delta method to the log of the CV$_B$ estimator.

To estimate the variance and bias for logit($\widehat{M}_1$), $\tau$ and $\beta$ may be replaced with respective estimates.   Variance estimates for the $\beta$ and $\tau$ estimators are available; e.g. the variance estimate for the DL estimator of $\tau$ can be computed from the REM and FEM weights and the $\tau$  estimate \citep[e.g.][]{biggerstaff1997incorporating} and variances for more complicated estimators of $\tau$ are attainable through packages such as \texttt{metafor} \citep{viechtbauer2010conducting}.  The variance estimate for the $\beta$ estimator is simply the inverse of the sum of unscaled weights (the inverse of the within-study variances) given as $$\widehat{\text{Var}}(\widehat{\beta})=\frac{1}{\sum^K_{i=1}W_i^*}$$
where $W_i^*=1/(v_i + \widehat{\tau}^2)$.  Note that the above variance is an estimate since $\widehat{\tau}$ replaces the unknown $\tau$.

\begin{corollary}\label{cor:var}
An approximate variance and bias for logit($\widehat{M}_2$) are
\begin{align*}
    \text{Var}[\text{logit}(\widehat{M}_2)] &= 4\times\text{Var}[\text{logit}(\widehat{M}_1)]\\
    \text{bias}[\text{logit}(\widehat{M}_2)] &= 2\times\text{bias}[\text{logit}(\widehat{M}_1)].
\end{align*}
\end{corollary}

\noindent We now consider some examples of the variance and bias of the $M_1$ and $M_2$ estimators.

\vspace{0.25cm}

\noindent \textbf{\textit{Example 1: variance based on the DL estimator}}

\vspace{0.25cm}

\noindent Recall that $W_i=1/v_i$ is the inverse variance weight for the $i$th study and is assumed fixed.  For simplicity we ignore truncation for the DL estimator in \eqref{T2}.  Under this assumption, \cite{biggerstaff1997incorporating} give the variance of the DL estimator of $\tau$ as
$$\text{Var}(T^2)=\frac{\text{Var}(Q)}{(S_1-S_2/S_1)^2}$$
where $\text{Var}(Q)=2(K-1) + c_1\tau^2 + c_2\tau^4$ for some constants $c_1$ and $c_2$ that depend on $S_1,S_2$ and $S_3$.  From Theorem \ref{thm:var}, unless $c_1=4(S_1-S_2/S_1)$ is close to zero, we can see that the first term in variance $\text{logit}(\widehat{M}_1)$ decreases with increasing $\tau^2$.  However, since the variance of $\widehat{\beta}$ increases with increasing $\tau^2$ as to whether the variance of $\text{logit}(\widehat{M}_1)$ increases or decrease with $\tau^2$ depends on the magnitude of $\beta$.  If $\beta$ is large, then the variance will decrease for some increasing $\tau^2$ before starting to increase as $\tau^2$ becomes large relative to $\beta^2$.

This example becomes simpler to illustrate if we assume that the within-study variances are small compared to $\tau^2$ (i.e. each $v_i<<\tau^2$).  Under this assumption, and the variance for $Q$ from, e.g., \cite{biggerstaff1997incorporating} \citep[who reference][]{larholt}, we have that
\begin{equation}\label{var_M1_smallv}
    \text{Var}[\text{logit}(\widehat{M}_1)]\approx \frac{1}{2}\left(S_2-2\frac{S_3}{S_1}+\frac{S_2^2}{S_1^2}\right) +  \frac{1}{K}\cdot \frac{\tau^2}{\beta^2}.
\end{equation}

Note that the first term in \eqref{var_M1_smallv} is fixed, depending on the within-study variances.  The second term is proportional to CV$_B^2$, indicating that the variance of the estimator increases as coefficient of variation increases.

\vspace{0.25cm}

\noindent \textbf{\textit{Example 2: bias based on the DL estimator}}

\vspace{0.25cm}

\noindent From the definitions seen in the previous example and Theorem \ref{thm:var}, we can see that the bias of of logit($\widehat{M_1}$) increases with $\tau^2$. As $\tau^2$ becomes large, the bias of the estimator depends on the magnitude of $\beta$ compared to $\tau^2$ and we see similar behaviour to the example above. However, if the magnitude of $\beta$ is large compared to $\tau^2$ (i.e. a small $\text{CV}_B)$, the bias of the estimator can also be large and negative.

In the case where $v_i << \tau^2$, we see that
\begin{equation}\label{biasM1_smallv}
    \text{bias}[\text{logit}(\widehat{M_1})] \approx \frac{1}{2}\left[\frac{1}{K}\frac{\tau^2}{\beta^2} - \left(S_2 - 2\frac{S_3}{S_1} + \frac{S_2^2}{S_1^2}\right)\right].
\end{equation}
Thus, as in the previous example, in the case where the within studies variances are small compared to $\tau^2$, the bias of the estimator increases with the coefficient of variation. 

\section{Confidence intervals}\label{sec:interval}

We now propose several potential interval estimators for CV$_B$, $M_1$ and $M_2$.  Some of these confidence intervals require a confidence intervals for $\tau^2$, which depend on the estimator chosen for $\tau$ and the interval estimator. In this paper we calculate the intervals for $\tau^2$ using the Q-profiling method \citep{viechtbauer2007confidence}. However, it is important to note that there are many ways of calculating these confidence intervals and for details see, e.g., \cite{viechtbauer2007confidence}. This paper also notes that coverages for bootstrap methods can vary wildly away from the nominal coverage. Thus, we will not consider bootstrap intervals in this paper.  

\subsection{Wald-type (WT) confidence intervals}

As noted in Section \ref{sec:point}, since $M_1$ and $M_2$ are bounded on $[0,\ 1]$ we consider intervals for the logit transformed statistics before back-transforming to the original scale.   Hence, our Wald interval for the logit of $M_i$ $(i=1,2)$ is
\begin{equation}
    [L_i, U_i] = \text{logit}(\widehat{M}_i)\pm z_{1-\alpha/2}\sqrt{\widehat{\text{Var}}[\text{logit}(\widehat{M}_i)]} 
\end{equation}
where $\widehat{\text{Var}}[\text{logit}(\widehat{M}_1)]$ and $\widehat{\text{Var}}[\text{logit}(\widehat{M}_2)]$ can be found in Theorem \ref{thm:var} and Corrolary \ref{cor:var} respectively.

This interval can then be back-transformed to the $M_i$ scale noting the transformation $\exp(\rho_i)/[1+\exp(\rho_i)]=\rho_i$. These logit transformed intervals often perform better than the regular Wald-type intervals.

Wald-type intervals for the $CV_B$ and log-transformed $CV_B$ measure were considered in \cite{takkouche2013confidence}. Given the link between CV$_B$, $M_1$ and $M_2$ given as
$$\log(\text{CV}_B)=\text{logit}(M_1)=\text{logit}(M_1)/2,$$
a Wald interval need only be found for one of the above, with the suitable back-transformation to return to the original scale for all three.

\subsection{Combining interval estimators for each parameter}\label{sec:combining}

In what follows let $[L_\tau(\alpha), U_\tau(\alpha)]$ and $[L_\beta(\alpha), U_\beta(\alpha)]$ denote $(1-\alpha)\times 100$\% confidence intervals for $\tau$ and $\beta$ respectively.  We also require intervals for $|\beta|$ and $\beta^2$ which requires special attention since, the signs of the lower and upper bounds differ for the interval for $\beta$, using the absolute or square transformation on the interval bounds does not result in an interval that contains zero, and will therefore have coverage lower than nominal.   We take a conservative approach to calculating the intervals that is based on transformation of the lower and upper bounds for the $\beta$ interval. 

As an example, we consider constructing an interval for $|\beta|$.  Based on the signs of the intervals for $\beta$, there are three scenarios:
\begin{enumerate}
\item If $L_\beta(\alpha)>0$ and $U_\beta(\alpha)>0$ then $L_{|\beta|}(\alpha)=|L_\beta(\alpha)|$ and $U_{|\beta|}(\alpha)=|U_\beta(\alpha)|$,

\item If $L_\beta(\alpha)<0$ and $U_\beta(\alpha)<0$ then $L_{|\beta|}(\alpha)=|U_\beta(\alpha)|$ and $U_{|\beta|}(\alpha)=|L_\beta(\alpha)|$,

\item If $L_\beta(\alpha)<0$ and $U_\beta(\alpha)>0$ then $L_{|\beta|}(\alpha)=0$ and $U_{|\beta|}(\alpha)=\text{max}\left(|L_\beta(\alpha)|,|U_\beta(\alpha)|\right)$.
\end{enumerate}

The conservativeness of this approach is associated with the third case above, where we choose the maximum of the transformed bounds as the upper bound.  An interval for $\beta^2$ can be similarly obtained by squaring the bounds of the interval for $\beta$.

In what follows, we discuss how to create some simple confidence intervals for $M_1$ and note that the generalisation to CV$_B$ and $M_2$ is similar in concept and therefore straightforward. Throughout we let $\widehat{\tau}$ and $\widehat{\beta}$ denote estimate of $\tau$ and $\beta$ where, like with the interval estimators, several different estimators are possible.  For simplicity, let $\text{CI}_{\text{CV}}(1-\alpha_\tau,1-\alpha_\beta)$, $\text{CI}_{M_i}(1-\alpha_\tau,1-\alpha_\beta)$ $(i=1,2)$ denote intervals for CV$_B$, $M_1$ and $M_2$ where the interval for $\tau$ is a $(1-\alpha_\tau)\times 100$\% confidence interval and for $\beta$ it is $(1-\alpha_\beta)\times 100$\%. 

\subsubsection{Confidence intervals with one fixed parameter}

When the confidence interval for one of the parameters is very narrow, then it may be possible to achieve good coverage by fixing that parameter, and simply using the lower and upper bounds of the other.  This approach has been used before; e.g. \cite{newcombe2011propagating} notes that \cite{khan1997seizure} used this approach to derive an interval for a two parameter function where the parameters were independently estimated.   In regards to the CV$_B$ measure of heterogeneity, \cite{takkouche2013confidence}, also used such an approach by fixing $\|\widehat{\beta}\|$ and using confidence intervals for $\tau$.  Using the notation above, fixing $\hat{\beta}$ is a $\text{CI}_{.}(0.95,0)$ interval and similarly $\text{CI}_{.}(0,0.95)$ when fixing $\widehat{\tau}$.

Choosing the placement of the lower and upper bounds depends on the variable, and whether the function defining the measure used is a decreasing or increasing function of the variable used.  For example, let $M_1(a,b)=a/(a + b)$ with both $a > 0$ and $b > 0$.  Then $\partial M_1(a,b)/(\partial a)=b/(a + b)^2 \geq 0$ so that our measure, $M_1$, is an increasing function of $\tau$.  Similarly, and more obviously, $M_1$ is decreasing in $\|\beta\|$.   
Then two possible intervals for $M_1$ are then
\begin{equation}\label{CIs:one_fixed}
    \text{CI}_{M_1}(0,.95)=\left[\frac{\widehat{\tau}}{\widehat{\tau}+U_{|\beta|}(\alpha)},\ \frac{\widehat{\tau}}{\widehat{\tau}+L_{|\beta|}(\alpha)}\right],\
    \text{CI}_{M_1}(.95,0)=\left[\frac{L_\tau(\alpha)}{L_\tau(\alpha)+\big|\widehat{\beta}\big|},\ \frac{U_\tau(\alpha)}{U_\tau(\alpha)+\big|\widehat{\beta}\big|}\right].
\end{equation}

Note that, and perhaps it is clearer, we can arrive at the same interval by writing $$1/M_1 = 1 + |\beta|/\tau.$$ Hence, using the idea above, a confidence interval for $1/M_1$ while fixing $|\widehat{\beta}|$ is
\begin{equation}
        \left[1 + \frac{|\widehat{\beta}|}{U_\tau(\alpha)},\ 1+\frac{|\widehat{\beta}|}{L_\tau(\alpha)}\right].
\end{equation}
A confidence interval for $M_1$ can then be found by inverting and switching the bounds of the interval for $1/M_1$ and we arrive back at \eqref{CIs:one_fixed}.

\subsubsection{Using both 95\% confidence intervals}

Another option to the above is to use the intervals simultaneously for both parameters.  As noted by \cite{newcombe2011propagating}, \cite{lloyd1990confidence} used this approach for an interval for the difference in correlated proportions which are typically conservative and preferable to those that too liberal.  

Then a confidence interval using the inversion approach above for $1/M_1$ while fixing $|\widehat{\beta}|$ is
\begin{equation}
        \text{CI}_{M_1}(0.95,0.95)\left[1 + \frac{L_{|\beta|}(\alpha)}{U_\tau(\alpha)},\ 1+\frac{U_{|\beta|}(\alpha)}{L_\tau(\alpha)}\right].
\end{equation}
A confidence interval for $M_1$ can then be found by inverting and switching the bounds of the interval for $1/M_1$.  As noted in the previous subsection, rearranging gives the confidence interval for $M_1$ as
\begin{equation}
        \left[\frac{L_\tau(\alpha)}{L_\tau(\alpha) + U_{|\beta|}(\alpha)},\ \frac{U_\tau(\alpha)}{U_\tau(\alpha) + L_{|\beta|}(\alpha)}\right].
\end{equation}

\subsubsection{Using two reduced-coverage intervals ($\alpha$-adjusted intervals)}\label{alph_adj}

To adjust for the conservative nature of the intervals above, another option is to decrease the coverage of the two intervals \citep{TryonW.W.2001Esde,TryonWarrenW2009EIPf,GoldsteinHarvey1995TGPo}. For a 95\% confidence interval, this means changing $\alpha$ from 0.05 to 0.1658. This method then calculates intervals based similar to the above with these adjusted value of $\alpha$.  These are therefore $\text{CI}_{\cdot}(0.8342,0.8342)$.

\subsubsection{PropImp}

Adjusting the $\alpha$ levels for the intervals used in the combined interval can work well in reducing the conservativeness associated with two 95\% confidence intervals.  However, adjusting the $\alpha$ to 0.1658 when, for e.g., one of the parameters is estimated precisely and one is highly variable, then increasing the $\alpha$ by this magnitude for the latter may result in liberal intervals.

Given the above, Propagating Imprecision \citep[PropImp][]{newcombe2011propagating} is an iterative approach of calculating confidence intervals where the $\alpha$ levels are adjusted differently, under constraints, for each of the lower and upper bounds, and to levels that maximise the interval width.  For a two parameter function for which a confidence interval is needed, PropImp can chose four different values for $\alpha$, one for each of the bounds for the intervals for each parameter. 

Define $f(X, Y)$ to be monotonic in both $X$ and $Y$, where, $(X_{z(\alpha)}^L, X_{z(\alpha)}^U)$ is a $(1 - \alpha)\times100\%$ confidence interval for $X$ and $z(\alpha)=\Phi^{-1}(1-\alpha/2)$ where $\Phi^{-1}$ is the standard normal inverse distribution function; e.g.  $\Phi^{-1}(0.975)\approx 1.96$. The confidence interval for $Y$,  $(Y_{z(\alpha)}^L, Y_{z(\alpha)}^U)$, is similarly defined. 

Then a confidence interval for $F$ can be formed from the confidence intervals for $X$ and $Y$ with three cases to consider.   These cases are
\begin{enumerate}
    \item $f(X, Y)$ is increasing in both $X$ and $Y$,
    \item $f(X, Y)$ is increasing in one and decreasing in the other,
    \item $f(X, Y)$ is decreasing in both.
\end{enumerate}

For CV$_B$, $M_1$ and $M_2$, we define $f$ such that $f(\tau, \beta)$ is equal to $\tau/|\beta|$, $\tau/(\tau + |\beta|)$ and $\tau^2/(\tau^2 + \beta^2)$ respectively.  Hence, for our measures we are dealing with Case 2 from above; increasing in $X$, decreasing in $Y$.   For Case 2,
we have intervals of the form ($L$, $U$), where
\begin{equation}
    L = \min_{0 \leq \theta_1 \leq \pi/2}f(X_{z\sin(\theta_1)}^L, Y_{z\cos(\theta_1)}^U), \\
    U = \max_{0 \leq \theta_2 \leq \pi/2}f(X_{z\sin(\theta_2)}^U, Y_{z\cos(\theta_2)}^L)
\end{equation}
noting that $\sin^2(\theta_1) + \cos^2(\theta_1)=1$.  Note the constraint arises since, e.g., for a 95\% confidence interval , $z=1.96$ and $\left[1.96\sin^2(\theta_1) + 1.96\cos^2(\theta_1)\right]=1.96^2$, and similarly with $\theta_2$.

Consequently, and as an example, the PropImp interval for CV$_B$ is
\begin{equation}\label{propimpint}
    L = \min_{0 \leq \theta_1 \leq \pi/2}\frac{\hat{\tau}_{z\sin(\theta_1)}^L}{|\hat{\beta}_{z\cos(\theta_1)}|^U},\ U = \max_{0 \leq \theta_2 \leq \pi/2}\frac{\hat{\tau}_{z\sin(\theta_2)}^U}{|\hat{\beta}_{z\cos(\theta_2)}|^L}
\end{equation}
where $\hat{\tau}_{z\sin(\theta_1)}^L$ and $\hat{\tau}_{z\sin(\theta_1)}^U$ are the lower and upper bounds of the interval for $\tau$, with similarly defined bounds for $\beta$.  As noted in \cite{newcombe2011propagating}, PropImp confidence intervals are generally wider then for other adjusted methods \citep{daly1998confidence}. However they tend to compare favourably when it comes to coverage.  To better understand PropImp in comparison to other intervals, we consider the following special cases:

\begin{description}
\item[Case 1: $\theta_1=\theta_2=\pi/4$.]  For this case we have $z\sin(\theta_1)=z\cos(\theta_1)=z\sin(\theta_2)=z\cos(\theta_2)=1.386$ and the resulting interval is the $\alpha$-adjusted interval from Section \ref{alph_adj}.  That is $\text{CI}_{\cdot}(0.8342,0.8342)$.
\item[Case 2: $\theta_1=\theta_2=\pi/2$.] Here, since we have $z\sin(\theta_1)=z\sin(\theta_2)=1.96$ and $z\cos(\theta_1)=z\cos(\theta_2)=0$, the resulting interval simply fixes $\widehat{\beta}$ in the lower and upper bounds and uses the interval for $\tau$ to determine the bounds.  Hence, $\text{CI}_{\cdot}(0.95,0)$.
\item[Case 3: $\theta_1=\theta_2=0$.]  Similar to Case 2, but the resulting interval is $\text{CI}_{\cdot}(0,0.95)$.
\end{description}

Hence, the domain of intervals considered for PropImp includes the aforementioned $\alpha$-adjusted interval and also the interval fixing one parameter, and using the interval for the other.  However, in practice other choices are likely to result.

\subsubsection{MOVER-R}

Extending interval estimators from \cite{DonnerAllan2012Ccif}, MOVER-R \citep{NewcombeRobertG2016Mcif} is a method for calculating confidence intervals for $\frac{\theta_1}{\theta_2}$, based on independent estimators $\widehat{\theta}_1$ and $\widehat{\theta}_2$. In exploring the use of these intervals, we found that MOVER-R was not in the context of CV$_B$, $M_1$ and $M_2$ since the lower bound for the interval for the denominator can be zero; e.g. the lower and upper bounds for the confidence interval for $\beta$ are of opposite sign.  As noted by \cite{NewcombeRobertG2016Mcif}, the PropImp method is more general, and given that it can be used in this scenario we do not consider MOVER-R in what follows.

\section{Simulations}

To assess the performance of our proposed intervals we performed simulation studies.  We start with conducting simulations focusing on the effect of varying $K$, $\tau$ and $\beta$. We then use settings similar to those for several real-data meta-analyses (e.g. sample sizes, $K$ etc.).   We found the bias-adjusting the Wald interval did not improve the performance of the intervals so the performance of the standard Wald intervals are reported.  Additionally, the combined 0.95 intervals were too conservative, and the use of one interval while fixing one parameter were too liberal.  We therefore have focused only on the $\alpha$-adjusted, PropImp and Wald intervals.   

\subsection{Simulated Data} 

We simulated standardised mean differences with equal sample sizes of 30 in each arm and varying the number of studies, $K = 10,\ldots,50$.  We also considered three values for $\beta$, $0.2$, $0.5$ and $0.8$ corresponding to small, moderate and large effects and values of $\tau$ set to $0.2, 0.4, 0.6$ and $0.8$. Observed effects were then randomly sampled from a non-central t-distribution under assumed normal distributions in each arm \citep[e.g. follow Proposition 2.1 of][adjusted to the two arms case]{malloy2013transforming}. We performed 10,000 trials for each possible setting. Note, due to the direct connections between CV$_B$, $M_1$ and $M_2$, the coverages for the intervals considered are the same and so that when reporting coverage, one result for each setting combination details coverage for each of the three measures.  Due to the large number of trials, we used the DerSimonian and Laird estimator of $\tau$ for computational efficiency.

For small $\tau$ and small number of studies, it is possible that $\widehat{\tau}=0$ due to truncation of the estimator to avoid negative estimates.  In practice, when this happens it is a pointless exercise to construct intervals for heterogeneity and confidence intervals for $\tau$ are similarly zero in both bounds.  Rather than remove these trials, we chose the largest interval, e.g. for $M_i$ $(0,\ 1)$ to indicate uncertainty for measuring heterogeneity. 

\begin{figure}[h!t]
    \centering
        \includegraphics[scale = 0.55]{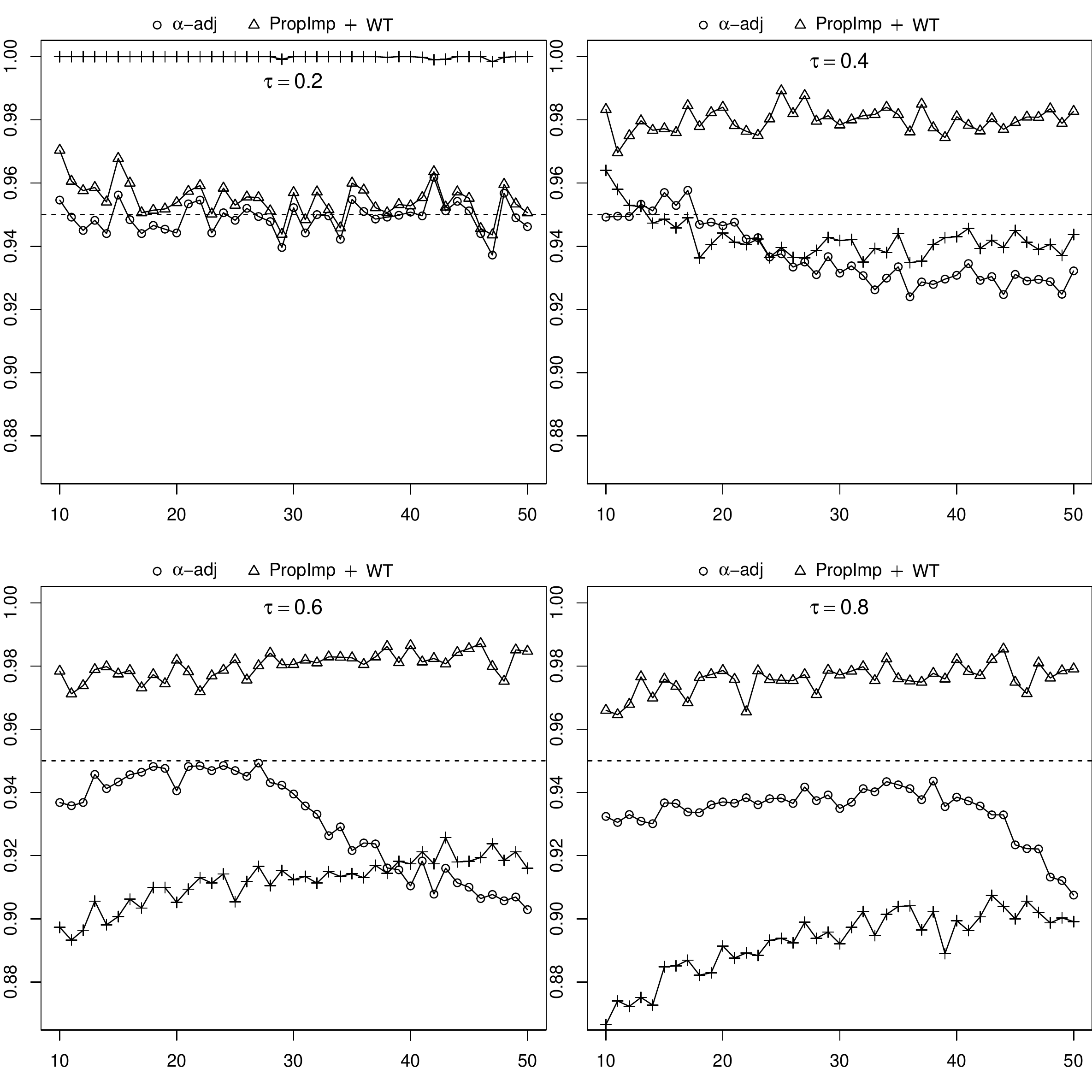}
    \caption{Line plots showing coverages for $\text{CV}_B, M_1$ and $M_2$ measures (note coverages for each setting for each measure are the same) on the vertical axis and the number of studies along the horizontal axis, for different values of $\tau$ when $\beta = 0.2$ and sample sizes of 30 in each study arm.}
    \label{fig:lines2}
\end{figure}

Figures \ref{fig:lines2} depicts coverages for $\beta=0.2$.  The Wald intervals were unreliable, being possibly very conservative otherwise, or too liberal and with coverage decreasing as heterogeneity increased.    The $\alpha$-adjusted intervals (i.e. CI$.(0.8342,0.8342)$ performed comparatively well, although for some settings were too liberal.  Given our belief that a conservative coverage is usually preferable, the PropImp intervals perform very well with coverages mostly between 0.95 and 0.98.

\begin{figure}[h!t]
    \centering
        \includegraphics[scale = 0.55]{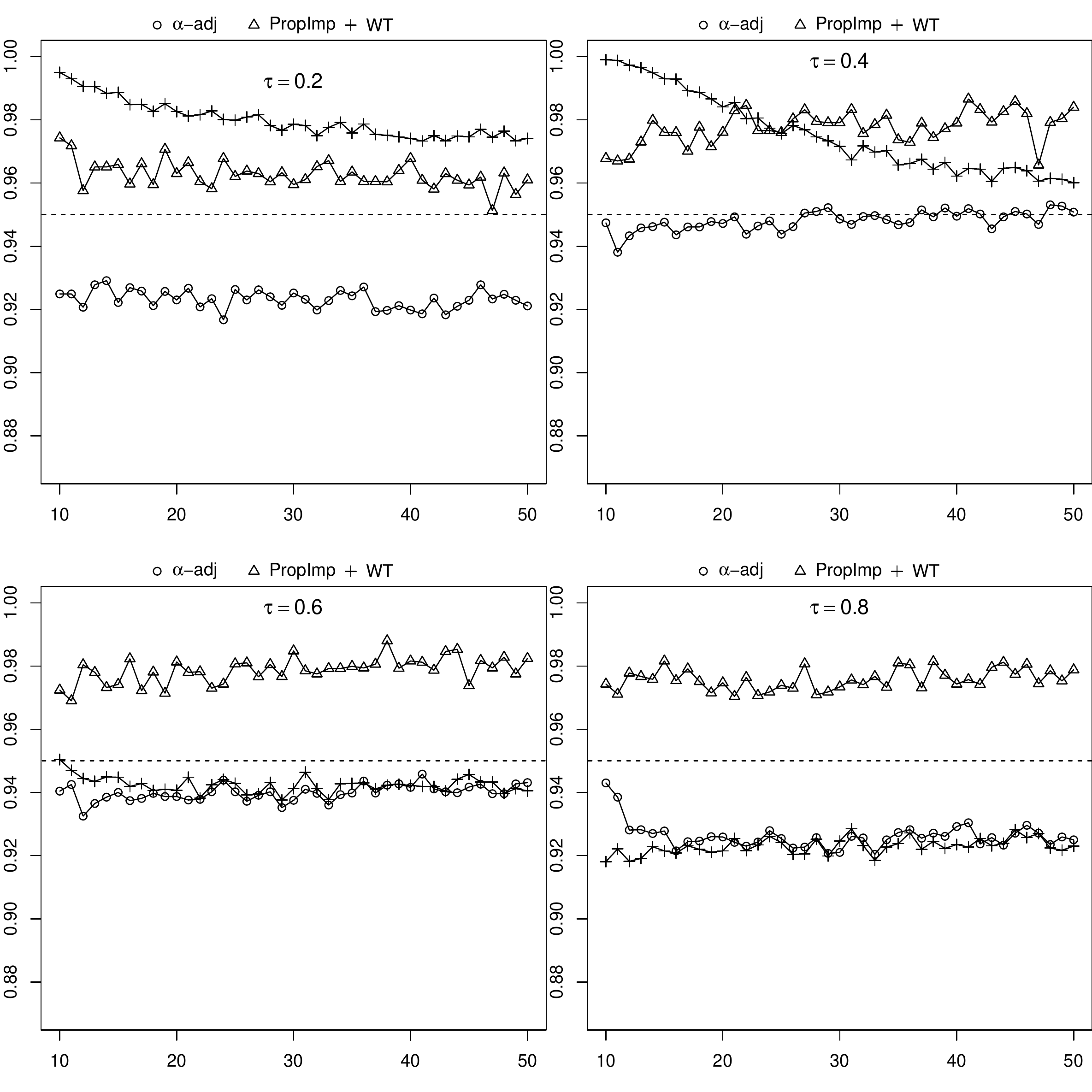}
    \caption{Line plots showing coverages for $CV_B, M_1$ and $M_2$ measures (note coverages for each setting for each measure are the same) on the vertical axis and the number of studies along the horizontal axis, for different values of $\tau$ when $\beta = 0.5$ and sample sizes of 30 in each study arm.}
    \label{fig:lines5}
\end{figure} 

\begin{figure}[h!t]
    \centering
        \includegraphics[scale = 0.55]{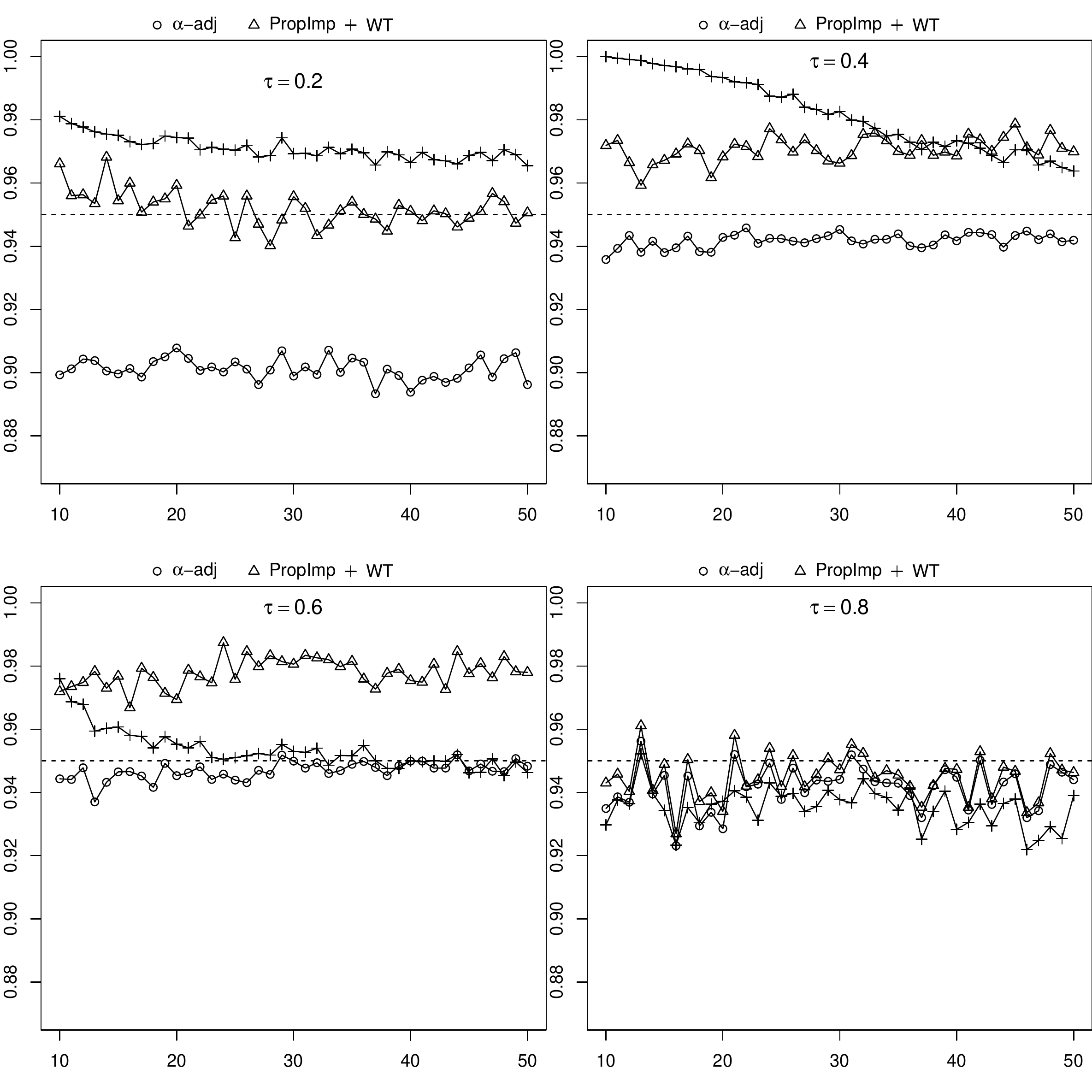}
    \caption{Line plots showing coverages for $CV_B, M_1$ and $M_2$ measures (note coverages for each setting for each measure are the same) on the vertical axis and the number of studies along the horizontal axis, for different values of $\tau$ when $\beta = 0.8$ and sample sizes of 30 in each study arm.}
    \label{fig:lines8}
\end{figure} 
\subsection{Simulations based on published data settings}

Simulation results for $\beta=0.5$ and 0.8 are shown in Figures \ref{fig:lines5} and \ref{fig:lines8}.  Results were similar although with typically better coverages for the Wald and $\alpha$-adjusted intervals.  The PropImp intervals were close to nominal or slightly conservative and therefore more reliable.

\begin{table}[h!t]
\caption{Settings based on three published meta-analysis to be used in the simulation studies.  The estimated values for CV$_B$, $M_1$ and $M_2$ are also provided.  Values of the within-study variances for \cite{zhu2020clinical} were calculated approximately from the report confidence intervals.}\label{table:sim_settings}
\centering
\begin{small}
\begin{tabular}{lll}
\toprule
Study &  settings: sample size ($N$) or within-study variances ($v$) \\ \midrule 
\cite{normand1999meta}   & $N = 311, 63, 146, 36, 21, 109, 67, 293, 112$ \\
\;\;$\widehat{\text{CV}}_B= 1.384$ &   (sample sizes split approximately equally between two arms)  \\ 
\;\;$\widehat{M}_1= 0.581$  &    \\ 
\;\;$\widehat{M}_2= 0.657$ &&   \\ \\

Bangert-Drowns  & $N = 60,\   34,\   95,\   209,\  182,\  462,\  38,\   542,\  99,\   77,\   40,\   190,\  113,\  50,\   47,\   44,$ \\
\;\;$\widehat{\text{CV}}_B=0.970$  & \;$24,\   78,\   46,\   64,\   57,\   68,\   40,\   68,\   48,\   107,\  58,\   225,\  446,\  77,\   243,\  39,\ 67$   \\ 
\;\;$\widehat{M}_1=0.492$  & \;$91,\   36,\   177,\  20,\   120,\  16,\   105,\  195,\  62,\   289,\  25,\   250,\  51,\   46,\   56$   \\ 
\;\;$\widehat{M}_2=0.485$  & (sample sizes split approximately equally between two arms)  \\ \\

\cite{zhu2020clinical}   & $v = 0.009,\  0.023,\  0.008,\  0.008,\  0.007,\  0.034,\  0.019,\  0.032,\  0.022,$ \\
\;\;$\widehat{\text{CV}}_B=0.227$  & \;\;\;\;\;\;\ $0.027,\  0.030,\   0.019,\  0.032,\  0.055,\  0.001,\  0.016,\  0.025,\  0.076,$   \\ 
\;\;$\widehat{M}_1=0.185$  & \;\;\;\;\;\;\ $0.023,\  0.013,\  0.020,\   0.036,\  0.010,\   0.007,\  0.022,\  0.028,\  0.023,$   \\ 
\;\;$\widehat{M}_2=0.049$  & \;\;\;\;\;\;\ $0.076,\  0.076,\  0.091,\  0.008,\  0.046,\  0.063,\  0.019,\  0.011 $  \\ 
 
 \bottomrule
\end{tabular}
\end{small}
\end{table}

We also considered simulations based on the settings for three published meta-analysis.  Summaries of the key settings, including the estimates of the measures, used are shown in Table \ref{table:sim_settings}.  The first two studies reporting standardized mean differences were the Hospital Stay of Stroke Patients data \citep{normand1999meta}, containing nine studies, and the Writing to Learn Interventions \citep{bangert2004effects} with 48 studies.  In both cases sample sizes ranged from small to large.  Both of these data sets are available in the \texttt{metafor} package \citep{viechtbauer2010conducting}.  To simulate data from the non-central t-distributions, we split the sample sizes approximately in the two arms.   Simulations based on \cite{zhu2020clinical}, which reported assumed approximately normal transformed incidence rates, generated data from the normal distribution using observed effects plus a generated random effect and within-study variances approximated from the reported intervals. These simulations were based on 10,000 simulated data sets and with $\beta$ set to the estimated value from the meta-analysis on the original data, $\tau$ was varied over $ 0.2, 0.4, 0.6, 0.8$. 

\begin{table}[h!t]
    \centering
        \caption{Coverages and weights for the three measures for data simulated according to the settings for different data sets (sample sizes and effect) but with varying $\tau$: $^*$median widths, $^1$\cite{normand1999meta},  $^2$\cite{bangert2004effects}, $^3$\cite{zhu2020clinical}}.
    \begin{tabular}{cccccc}
    \hline
      settings & measure & coverage & width (CV$_B$) & width ($M_1$)& width ($M_2$)  \\
      \hline
      $\tau=0.2$ \\
         HSSP$^1$ $(\beta = 0.537)$ & $\alpha_{adj}$ & 0.920 & 0.910$^*$ & 0.403 & 0.466\\
          & PropImp & 0.963 & 1.123$^*$ & 0.478 & 0.548\\
         & WT & $0.997$ & 1.057$^*$ & 0.553 & 0.605\\
       WLI$^2$ $(\beta=0.222)$   & $\alpha_{adj}$ & 0.948 & 1.224$^*$ & 0.310 & 0.524\\
       & PropImp & 0.978 & 1.604$^*$ & 0.380 & 0.612\\
         & WT & 0.989 & 1.131$^*$ & 0.300 & 0.519\\
       Zhu$^3$ $(\beta =2.225)$   & $\alpha_{adj}$ & 0.868 & 0.053 & 0.045 & 0.011\\
         & PropImp & 0.950 & 0.071 & 0.060 & 0.013\\
         & WT & 0.970 & 0.091 & 0.076 & 0.024\\
         $\tau = 0.4$ \\
          HSSP$^1$ $(\beta = 0.537)$    & $\alpha_{adj}$ & 0.937 &1.720$^*$ & 0.434 & 0.646\\
          & PropImp & 0.967 & 2.494$^*$ & 0.525 & 0.734\\
         & WT & $0.988$ & 1.479$^*$ & 0.446& 0.662\\
       WLI$^2$ $(\beta=0.222)$   & $\alpha_{adj}$ & 0.928 & 2.794$^*$ & 0.282 & 0.388\\
        & PropImp & 0.979 & 4.512$^*$ & 0.361 & 0.475\\
         & WT & 0.942 & 2.543$^*$ & 0.314 & 0.474\\
       Zhu$^3$ $(\beta =2.225)$   & $\alpha_{adj}$ & 0.900 & 0.089 & 0.065 & 0.038\\
         & PropImp & 0.951 & 0.107 & 0.076 & 0.038\\
         & WT & 0.967 & 0.168 & 0.118 & 0.062\\
    $\tau=0.6$ \\
         HSSP$^1$ $(\beta = 0.537)$     & $\alpha_{adj}$ & 0.932 &3.331$^*$ & 0.458 & 0.662\\
          & PropImp & 0.970 & 6.709$^*$  & 0.552 & 0.751\\
         & WT & 0.933 & 2.417$^*$ & 0.491 & 0.733\\
       WLI$^2$ $(\beta=0.222)$   & $\alpha_{adj}$ & 0.898 & 6.270$^*$ & 0.281 & 0.282\\
       & PropImp & 0.973 & 15.853$^*$ & 0.358 & 0.358\\
         & WT & 0.924 & 5.105$^*$ & 0.375 & 0.476\\
       Zhu$^3$ $(\beta =2.225)$   & $\alpha_{adj}$ & 0.916 & 0.137 & 0.085 & 0.067\\
     & PropImp & 0.956 & 0.158 & 0.096 & 0.078\\
         & WT & 0.971 & 0.252 & 0.151 & 0.125\\
    $\tau=0.8$ \\
      HSSP$^1$ $(\beta = 0.537)$        & $\alpha_{adj}$ & 0.935 &6.657$^*$ & 0.467 & 0.626\\
          & PropImp & 0.972 & 45.421$^*$ & 0.558 & 0.719\\
         & WT & 0.912 & 3.876$^*$ & 0.549 & 0.781\\
    WLI$^2$ $(\beta=0.222)$      & $\alpha_{adj}$ & 0.894 & 14.061$^*$ & 0.273 & 0.221\\
    & PropImp & 0.978 & $\infty^*$ & 0.344 & 0.288\\
         & WT & 0.904 & 8.796$^*$ & 0.434 & 0.509\\
      Zhu$^3$ $(\beta =2.225)$    & $\alpha_{adj}$ & 0.933 & 0.198 & 0.102 & 0.111\\
        & PropImp & 0.966 & 0.229 & 0.118 & 0.131\\
         & WT & 0.970 & 0.342 & 0.176 & 0.191\\
    \hline
    \end{tabular}
    \label{tab:my_label}
\end{table}

Coverage probabilities for each of the variables can be found in Table \ref{tab:my_label}.  The performance of the Wald intervals was at times too conservative, although in many cases the coverage was very good.  The Wald-type intervals for CV$_B$ were obtained exponentiating the Wald-type interval for the logit transformed $M_1$ (recalling the link between the measures).  A disadvantage of this Wald interval for CV$_B$ is that following back-transformation the upper bound could be extremely large.  This results in meaningless average widths (e.g. not representing typical width) and so the medians were reported.  

The $\alpha$-adjusted method typically displayed reasonable coverage although tended to be liberal, and for the Zhu study setting coverage dropped to just 0.868 for small $\tau$. Overall, the PropImp intervals, had excellent coverage, being close to the nominal 0.95 for all settings. 

The widths were often large for the CV$_B$ measure, in particular when $\beta$ is small relative to $\tau$.  However, by reporting on the $[0,\ 1]$ scale like for $M_1$, we avoid the problem of very large widths and there the intervals may be easier to interpret.

\section{Examples}
Let us now return to our motivating example where we considered a meta-analysis containing 35 studies and using a single armed, random-effects analysis of double arcsin transformed incidence rates.  Due to the good coverages obtaqined in the simulations, we report the PropImp intervals. Calculating $CV_B$ and its confidence interval, we have $CV_B = 0.227$  and $(0.199, 0.259)$. For the other measures, we have $M_1 = 0.185$ $(0.166, 0.206)$ or $M_2 = 0.049$ $(0.038, 0.063)$. Hence, the researcher can state that there does appear to be heterogeneity present in the analysis although the amount of heterogeneity should be considered small relative to the effect being estimated.

\begin{table}[h!t]
    \centering
        \caption{Estimates and confidence intervals for the three examples. We display $95\%$ confidence intervals for the $\alpha$-adjusted and PropImp methods for CV$_B$ and M$_1$ measures.}
    \begin{tabular}{ccccc}
    \hline
      Dataset & Measure &  Estimate & $\alpha_{adj}$ & PropImp  \\
      \hline
      HSSP   & $I^2$ & 93.534$\%$ &  &  \\ 
    $(\widehat{\tau}^2 = 0.540)$ & CV$_B$ & 1.384 & (0.733, 8.358) & (0.624, 57.574)\\ 
      &M$_1$ & 0.581 & (0.423, 0.893)  & (0.384, 0.983)\\ 
      WLI   & $I^2$ & 56.118$\%$ &  & \\ 
    $(\widehat{\tau}^2 = 0.046)$ & CV$_B$ & 0.970 & (0.685, 2.223) & (0.624, 2.665)\\ 
      &M$_1$ & 0.492& (0.407, 0.690)  & (0.384, 0.727)\\
      Zhu   & $I^2$ & 95.01\% &  & \\ 
    $(\widehat{\tau}^2 = 0.255)$ & CV$_B$ & 0.227 & (0.193, 0.267) & (0.199, 0.259)\\ 
      &M$_1$ & 0.185 & (0.162, 0.211) & (0.166, 0.206)\\ 
\hline
    \end{tabular}
    \label{tab:examples}
\end{table}

Further examples can be found in Table \ref{tab:examples}, where we provide estimates and confidence intervals for each of the data sets whose settings were used to conduct simulations in the previous section.  Despite the similar values for $I^2$ for the second and third data sets, we see that the estimates, and intervals, for CV$_B$ and $M_1$ are very different with either large or small relative heterogeneity.

\section{Conclusion}

In this paper we propose new intervals for the coefficient of variation used to assess heterogeneity in meta-analyses. We also suggest to transform the measure onto a $[0,\ 1]$ scale, to avoid the possibility of very large values of the measure and associated confidence intervals. We discuss similarities between the transformed measures and highlight that they can be used as a population estimate of heterogeneity. We suggest that of the two measures we propose, $M_1$ would be preferred due to it being a simpler re-scaling of the coefficient of variation.   There are two key relationships here that are relevant being
$$\text{logit}(M_1)=\log(\text{CV}_b),\; M_1^{-1}=1+1/\text{CV}_B$$
meaning that it is simple to covert point and interval estimators from the $M_1$ scale to the CV$_B$ scale, and vice versa.

Several confidence intervals for the coefficient of variation and the transformed measures were considered that exhibit good coverage properties.  We recommend the slightly conservative PropImp intervals \citep{newcombe2011propagating} which can be reliably used in practice.  Wald type intervals had varying performance.  A simple alternative to the PropImp intervals is to combine reduced coverage intervals (coverage of each set to 0.8342) since this interval typically had reasonable coverage.

\section*{Supplementary Materials}
All simulation code and data is available at \url{https://osf.io/yq65h/}

\clearpage

\appendix

\section{Approximate variance and bias for the logit transformed $M_1$ estimator}

Recall that $\text{logit}(\widehat{M}_1)=\log(\widehat{\text{CV}}_B)=\log(\widehat{\tau})-\log(|\widehat{\beta}|)$ so that $\text{logit}(\widehat{M}_1)=\log(\sqrt{v^2})-\log(|\widehat{\beta}|)$ where $v=\widehat{\tau}^2$.  Using the Delta method, the second order Taylor Series expansion gives
\begin{align*}
    \text{logit}(\widehat{M}_1) \approx &\text{logit}(M_1) + (\widehat{\tau}^2-\tau^2)\cdot \frac{1}{2\tau^2} - (\widehat{\beta}-\beta)\cdot \frac{\text{sign}(\beta)}{|\beta|}\\
    &+\frac{1}{2}\left[(\widehat{\beta}-\beta)^2\frac{1}{\beta^2}-(\widehat{\tau}^2-\tau^2)^2\frac{1}{2\tau^4}\right].\label{2ndordM1}
\end{align*}
where $\text{sign}(x)=1$ if $x>0$ and $\text{sign}(x)=-1$ if $x<0$.  Note this is not defined for when $\beta=0$ although nor is the CV$_B$.  Hence, a first order approximate variance, assuming that $\widehat{\tau}$ and $\widehat{\beta}$ are uncorrelated, is equal to
\begin{equation}
    \text{Var}\left[\text{logit}(\widehat{M}_1)\right]\approx \text{Var}(\widehat{\tau}^2)\cdot \frac{1}{4\tau^4} + \text{Var}(\widehat{\beta})\cdot \frac{1}{\beta^2}.\label{VarM1}
\end{equation}

An approximation to the bias can be found by taking the expected value of the second order term and is therefore equal to
\begin{equation}
\frac{1}{2}\left[\text{Var}(\widehat{\beta})\cdot\frac{1}{\beta^2}-\text{Var}(\widehat{\tau}^2)\cdot\frac{1}{2\tau^4}\right]\label{biasM1}
\end{equation}


\newpage
\
\bibliographystyle{authordate4}
\bibliography{main}

\end{document}